\documentclass[12pt,preprint]{aastex}

\begin{document}

\title{Spin-Induced Disk Precession in Sagittarius A*}

\author{Gabriel Rockefeller\altaffilmark{1,2}, Christopher L.
  Fryer\altaffilmark{1,2}, and Fulvio Melia\altaffilmark{1,3,4}}
\altaffiltext{1}{Department of Physics, The University of Arizona,
  Tucson, AZ 85721}
\altaffiltext{2}{Theoretical Division, Los Alamos National Laboratory,
  Los Alamos, NM 87545} 
\altaffiltext{3}{Steward Observatory, The University of Arizona,
  Tucson, AZ 85721}
\altaffiltext{4}{Sir Thomas Lyle Fellow and Miegunyah Fellow}

\begin{abstract}
  
  In Sgr~A* at the Galactic center, by far the closest and easiest
  supermassive black hole we can study, the observational evidence is
  increasingly pointing to the presence of a compact, hot, magnetized
  disk feeding the accretor.  In such low-Mach-number plasmas, forces
  arising, e.g., from pressure gradients in the plasma, can altogether
  negate the warping of disks around Kerr black holes caused by the
  Bardeen-Petterson effect and can lead to coherent precession of the
  entire disk.  In this {\it Letter}, we present for the first time
  highly detailed 3D SPH simulations of the accretion disk evolution
  in Sgr A*, guided by observational constraints on its physical
  characteristics, and conclude that indeed the Bardeen-Petterson
  effect is probably absent in this source. Given what we now
  understand regarding the emission geometry in this object, we suggest 
  that a $\sim 50-500$-day modulation in Sgr~A*'s spectrum, arising
  from the disk precession, could be an important observational
  signature; perhaps the $\sim 106$-day period seen earlier in its
  radio flux, if confirmed, could be due to this process.  On the
  other hand, if future observations do not confirm this long 
  modulation in Sgr A*'s spectrum, this would be an indication that 
  either the disk size or orientation is very different from current 
  estimates, or that the black hole is not spinning at all (unlikely), 
  or that our current understanding of how it produces its radiative 
  output is incorrect.
\end{abstract}


\section{Introduction}
At the Galactic center, the compact radio source Sgr~A* appears to be
the radiative manifestation of an accreting $3.7\times 10^6\, M_\odot$
supermassive black hole \citep{Ghez05}.  Its radio to mm spectrum is
thought to be a composite of two principal components---a slightly
inverted power-law with a notable bump at mm/sub-mm wavelengths.  The
latter is somewhat reminiscent of the ``big blue bump'' seen in many
Active Galactic Nuclei (AGNs), evidently produced by an accretion disk
feeding the supermassive black hole \citep[see, e.g.,][]{B01}.  But
unlike the blackbody emission process that gives rise to the
optical/UV thermal feature in an AGN spectrum, the radiative mechanism
responsible for Sgr~A*'s mm/sub-mm spectral excess appears instead to
be a combination of thermal and non-thermal synchrotron emission
within the inner portion ($r < 10\, r_S$) of a hot, magnetized
Keplerian flow \citep{MLC01,LM01}.  (The Schwarzschild radius, $r_S$,
for an object of this mass $M$ is $2GM/c^2\approx 10^{12}\, \rm{cm}$,
or roughly $1/15$~A.U.)  The inferred characteristics of the compact
region surrounding Sgr~A* are also consistent with the $\sim 10\%$
linear polarization detected from this source at mm wavelengths
\citep{MLC00,BML01}.  Given the complex spatial arrangement of the
mass-losing stars that feed this object, it is unlikely that the disk
angular momentum axis is aligned with the spin axis of the black hole.

The possible (but currently unconfirmed) detection of a 106-day cycle
in Sgr~A*'s radio variability \citep{Z01} has added significant
intrigue to this picture, since it may indicate precession of the disk
induced by the spin of the black hole \citep{LM02b}.  The dynamical
time scale near the marginally stable orbit around an object with this
mass is $\approx 20$ minutes.  Thus, since the physical conditions
associated with the disk around Sgr~A* imply unwarped, coherent
precession, a precession period of 106 days may be indicative of a
small black-hole spin ($a \sim 0.1$) if the circularized flow is
confined to a region within $\sim 30\,r_S$.  (Throughout this paper,
we define $a$ to be the black hole angular momentum divided by
$GM^2/c$.)  The precession of a larger structure at the same rate
would require a bigger black hole spin; alternatively, for a given
black hole spin, a larger disk would precess with a longer period.
\citet{LM02b} also noted that a small value of $a$ ($< 0.1$) would be
favored if the non-thermal ($\sim 1-20$ cm) portion of Sgr A*'s
spectrum is powered with energy extracted via a Blandford-Znajek type
of process, for which the observed luminosity would correspond to an
outer disk radius $r_{out}\sim 30\, r_S$.  In addition, a small disk
size was suggested by earlier hydrodynamical and magnetohydrodynamical
simulations \citep{CM97,IN02}, and is implied by Sgr A*'s spectral and
polarimetric characteristics.

For a thin, cold disk, the differential Lense-Thirring precession
dominates the internal coupling of the plasma at small radii and
therefore leads to the so-called Bardeen-Petterson effect
\citep{BP75}, in which the inner region flattens toward the equatorial
plane, producing a warped accretion pattern.  As shown by \citet{NP00}
and confirmed recently by \citet{FA05}, however, thicker disks with a
mid-plane Mach number of $\sim 5$ or less can suppress warping effects
due to the coupling provided by pressure gradients and viscosity in
the gas.  The mid-plane Mach number in Sgr~A* is $\sim 3$ \citep[for
detailed comparisons of model disks to observations, see,
e.g.,][]{MLC01,IN02}, so it seems that the disk in this system may be
precessing as one unwarped structure \citep{LM02b}.  In this paper, we
report the results of smoothed particle hydrodynamics (SPH)
simulations we have carried out to test these ideas at a higher level
of sophistication than was attempted earlier by \citet{LM02b}.


\section{Physical Principles and Method}\label{sect:principles}
We evolve gaseous disks in the gravitational potential of the black
hole at the Galactic center using the three-dimensional smoothed
particle hydrodynamics (SPH) code described in \citet{F05}.  The inner
boundary of the calculation is placed at a radius corresponding to the
innermost stable circular orbit, as a function of $a$.  For simplicity
in this first generation of calculations, particles crossing the inner
boundary carry all of their mass and angular momentum into the black
hole, and no additional infalling material enters the simulation at
the outer edge of the disk; we also do not model the magnetic field or
evolve the equations of magnetohydrodynamics, but we include a
prescription for the role of the magnetorotational instability in
transporting angular momentum (see below).  In principle, infalling
material can exert an additional torque on the disk with a tendency to
align the disk's angular momentum vector with that of the accreting
plasma.  However, the angular momentum flow through the disk is also
subject to magnetic coupling effects at its inner boundary
\citep{KH02}, which can either offset or enhance the effect of the
external torque, depending on the configuration of the magnetic field
\citep[see][]{LM02b}.  We do not fully explore the range of inner and
outer boundary conditions in this first treatment and instead leave
this important survey to future publications.  Consequently, our
simulated disks mediate an outward viscous transport of angular
momentum, which causes them to slowly grow in radius.

We construct disks based on current estimates of flow properties in
the Galactic center \citep{MLC01}, and we incorporate important
effects of general relativity in a Kerr metric and an anomalous
viscosity due to the magnetorotational instability that the
observations strongly indicate is active in this system.  The
gravomagnetic force per unit mass and the anomalous viscosity dominate
the dynamics of the disk; even at its turbulent saturation point, the
magnetic field energy represents only about $15$ percent of the
equipartition value.  Since the adoption of a post-Newtonian approach
to handle the general relativistic effects (see below) introduces
$\sim 10$ percent errors, it is reasonable to ignore the dynamical
impact of the magnetic field for the first set of simulations.
Advancing this study to the next level of sophistication will entail
the use of a more elaborate magnetohydrodynamic treatment, which is
currently in development.  In the final analysis, we will indeed
follow the evolution of the inner disk in Sgr A* with all effects
included in the simulation: the magnetic field and associated
anomalous viscosity and the gravomagnetic force per unit mass.  The
calculations we describe here will provide a powerful baseline against
which the later calculations may be compared and tested.

\subsection{The Conservation Equations and SPH Code}
Since the gravitational influence of interest simulated here occurs
many Schwarzschild radii from the black hole, we follow \citet{NP00}
and include the relativistic effects only in a post-Newtonian
approximation.  Our momentum equation therefore takes the form
\begin{equation}
\frac{d\vec{v}}{dt} = -\frac{1}{\rho}\vec\nabla P + \vec{v}\times\vec{h}
-\vec\nabla\Phi+\vec{F}_{visc}\;,
\end{equation}
where $\rho$ is the density, $\vec{v}$ is the velocity, $P$ is the
pressure, $\Phi$ is the gravitational potential, and $\vec{F}_{visc}$
is the viscous force per unit mass.  The term $\vec{v}\times\vec{h}$
is the lowest-order post-Newtonian approximation to the gravomagnetic
force per unit mass near a rotating black hole \citep[see,
e.g.,][]{B96}.  In this equation, $\vec{h}$ is defined as
\begin{equation}
\vec{h}\equiv {2\vec{S}\over r^3}-{6(\vec{S}\cdot\vec{r})\vec{r}\over r^5}\;,
\end{equation}
where
\begin{equation}
\vec{S}={G\vec{J}\over c^2}
\end{equation}
in terms of the (cylindrical) coordinate vector
$\vec{r}=(\eta,\phi,z)$, and $r=|\vec{r}|$.  The spin angular momentum
of a Kerr black hole (using our definition of $a$; see above) is given by
\begin{equation}
\vec{J}={aGM^2\over c}\,\hat{k}\;,
\end{equation}
where $\hat{k}$ denotes the unit vector in the $z$-direction.  We
calculate the gravitational acceleration due to the black hole using
\begin{equation}
-\vec\nabla\Phi = -\frac{GM}{r^3}\left(1 + \frac{6r_{g}}{r}\right)\vec{r}
\end{equation}
(where $r_g \equiv r_S/2= GM/c^2$), which produces the correct apsidal
precession frequency at large distances from the black hole
\citep{NP00}.  

We use an equation of state of the form $P = K\rho^{\gamma}$---where
$P$ is the pressure, $\gamma = 5/3$ is the heat capacity ratio, and
$K$ is a constant chosen to set the midplane Mach number.  Energy
dissipated through artificial viscosity is allowed to leave the
system.

\subsection{The Anomalous Viscosity}

Many of the properties derived by \citet{BH91,BH92} for weakly
magnetized accretion disks appear to be present in Sgr~A*.  One of the
most important defining characteristics is the existence of an
anomalous viscosity arising from the Maxwell stress, which in Sgr~A*
easily dominates over the Reynolds stress.  In our calculations, we
model the effect of this anomalous viscosity using the $\alpha$-disk
prescription
\begin{equation}
\nu=\frac{\alpha c_s^2}{\Omega},
\end{equation}
where $c_s$ is the local sound speed and $\Omega$ is the Keplerian
angular velocity.  Earlier quasi-analytical fits to Sgr~A*'s spectrum,
and analysis of X-ray flares detected from this object by
\textit{Chandra} \citep{LM02a} concluded that the ratio $\beta_0$ of Maxwell
stress to thermal pressure is $\simeq 0.05$; since the viscosity
$\nu$ can also be written as
\begin{equation}
\nu=\frac{2}{3}\frac{\beta_0 c_s^2}{\Omega},
\end{equation}
we relate $\alpha$ and $\beta_0$ using $\alpha = (2/3) \beta_0$ and find
that $\alpha$ should be approximately 0.03 in our model disks.

We implement the viscosity using the same technique used by
\citet{NP00}, i.e., we adjust the bulk viscosity coefficient
$\alpha_{SPH}$ and the Von Neumann-Richtmyer viscosity coefficient
$\beta_{SPH}$ of the standard SPH artificial viscosity prescription.
Following \citet{NP00}, we set $\alpha_{SPH} = 0.5$ and $\beta_{SPH} =
0.0$.  By tracking the motion of particles in our calculations,
measuring the ratio of orbital period $P$ to accretion timescale
$\tau_{acc}$, and relating the ratio of timescales to the viscosity
parameter $\alpha \simeq P/\tau_{acc}$, we find that a value of
$\alpha_{SPH} = 0.5$ corresponds to a value of $\alpha \simeq 0.02$.

\section{Results}\label{sect:results}
The characteristics of the three simulations we ran that directly
address the questions we wish to answer in this paper are summarized in
Table~\ref{sims}.  Simulation E1 is a direct comparison with the
calculation of the same name reported by \citet{NP00}.  Simulation GC20
was constructed with an initial outer radius of $20\, r_S$; the
parameters of this model were chosen to be fully consistent with the
physical conditions outlined above.  It is the contrast between this
model and E1 that we expect to highlight the difference between the
behavior of a cold disk that is subject to the Bardeen-Petterson
effect and the low-Mach-number disk that we believe is present at the
Galactic center; according to earlier semi-analytical analysis, the
Galactic center disk ought not to experience this effect.  However,
because we have chosen not to feed the disk from outside, the viscous
transport of angular momentum leads to a slow growth in its size; to
test the dependence of the disk's temporal behavior on the initial
conditions, we also carried out simulation GC30, which has an initial
size of $30\, r_S$ and is otherwise identical to GC20.

\begin{figure}
\epsscale{1.00}
\plotone{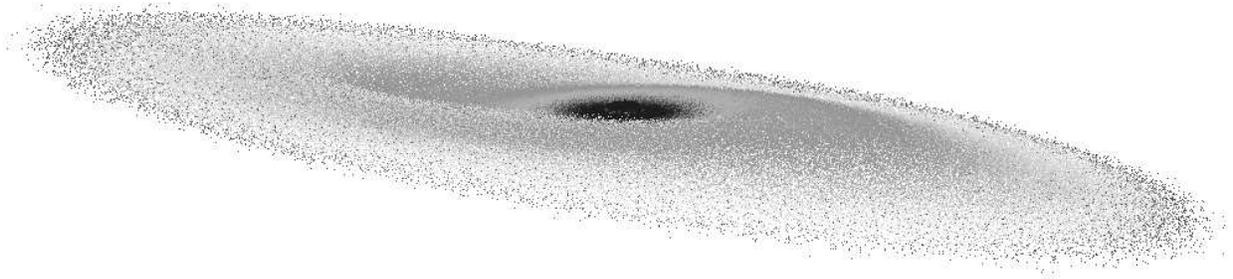}
\caption{The 3-D arrangement of $326,034$ SPH particles in a reproduction 
  of test E1 of \citet{NP00}.  The inner portion of the inclined disk
  is warped into the equatorial plane of the black hole (aligned
  horizontally in the image) through the action of the
  Bardeen-Petterson effect.}
\label{fig:3D-NP}
\end{figure}

Figure~\ref{fig:3D-NP} shows the 3-D arrangement of SPH particles in
our reproduction of test E1 by \citet{NP00}.  In this image, the black
hole spin axis points in the vertical direction.  The outer portion of
the disk has maintained its original 10-degree inclination relative to
this direction.  However, due to the Bardeen-Petterson effect, the
inner portion of the disk has become warped and now lies in the
equatorial plane of the black hole.  By this time in the simulation,
which corresponds to 4 Keplerian orbital periods at $r = 25\, r_S$,
the transition region between the outer inclined portion of the disk
and the inner warped portion has stabilized at a radius of $7\, r_S$,
in good agreement with \citet{NP00}.

\begin{figure}
\epsscale{1.00}
\plotone{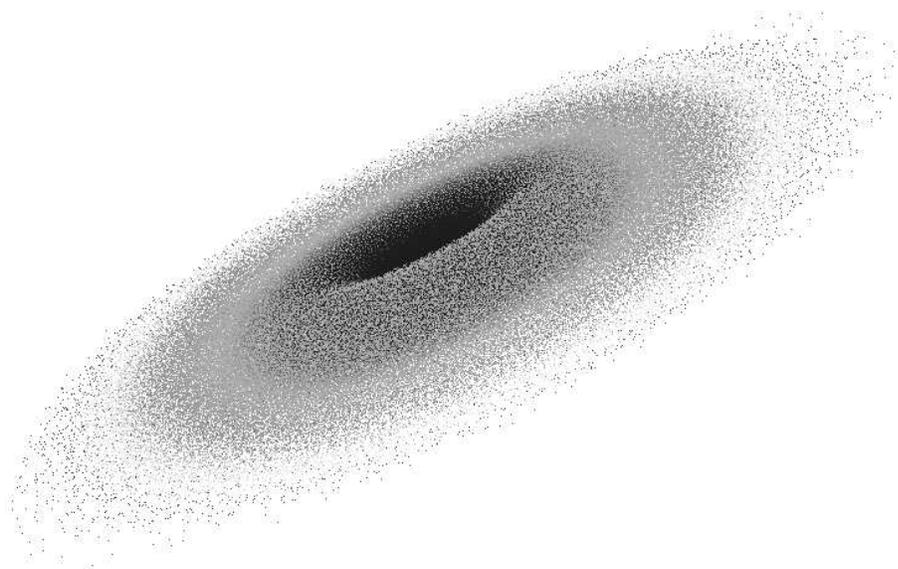}
\caption{The 3-D arrangement of $423,626$ SPH particles in a simulated 
  disk constructed using the gas temperature and density, constrained
  by the observations, in the environment near the black hole at the
  Galactic center.  The disk is relatively thick, so the
  Bardeen-Petterson effect is suppressed; even the innermost portion
  of the disk remains aligned with the outer disk and is not warped
  into the equatorial plane of the black hole.}
\label{fig:3D-GC}
\end{figure}

In contrast, a disk constructed according to the best estimates for
conditions in the Galactic center does not warp; instead, the entire
disk remains tilted out of the equatorial plane and precesses around
the spin axis of the black hole.  Figure~\ref{fig:3D-GC} shows the 3-D
arrangement of SPH particles in simulation GC30 after 85 Keplerian
orbits at $30\, r_S$, when the disk has an outer radius of $55\, r_S$;
the central portion of the disk lies in the same plane as the rest of
the disk, and not in the equatorial plane of the black hole.
\citet{NP00} also considered one case in which the midplane Mach
number was low (specifically, $\mathcal{M} = 5$) and found that that
particular disk did not warp.  Our result for model GC30, which has a
midplane Mach number of $3$, is consistent with their finding.
Simulation GC20 behaves in exactly the same way.  A principal result
of this work is that the physical conditions inferred from Sgr~A*'s
spectrum evidently imply that the entire disk precesses coherently
around the black hole spin axis, confirming the prediction of
\citet{LM02b}.

An important observational signature of this effect is the dependence
of the precession period on the size of the disk for a given value of
the black hole spin parameter.  The size of the disk provides a
measure of the moment of inertia of the structure, whose response to
the applied gravomagnetic torque determines the rate of precession.

\begin{figure}
\epsscale{1.00}
\plotone{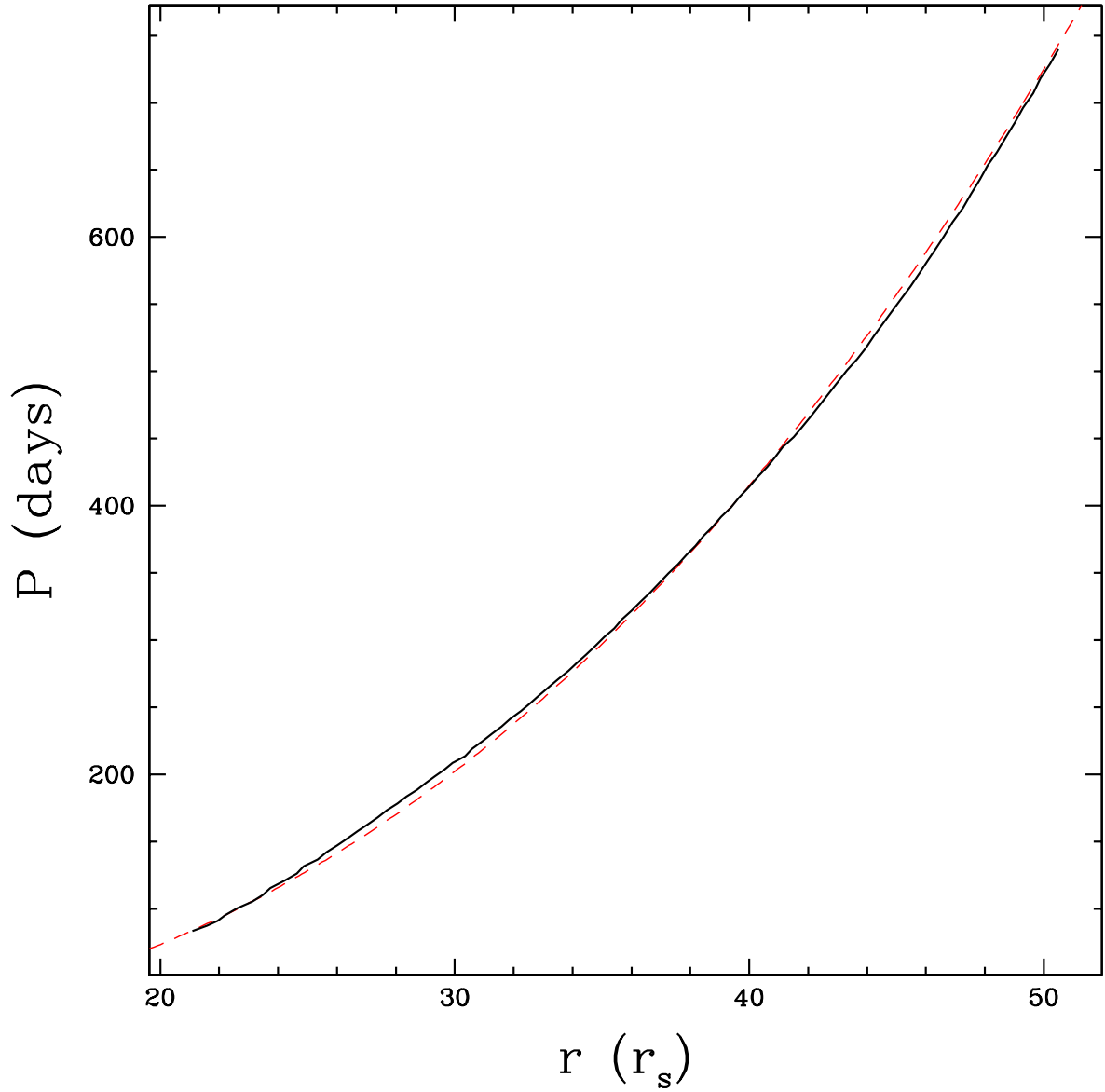}
\caption{The precession period of the simulated Galactic center disk 
  versus the size of the disk.  The solid curve, measured from the
  simulation, matches the $P \propto r^{5/2}$ relationship (shown as a
  thin dashed line) predicted by \citet{LM02b}, demonstrating that a
  disk formed in the Galactic center likely does precess as one
  coherent, unwarped structure.}
\label{fig:pvsr}
\end{figure}

\citet{LM02b} argued that the precession period of a
coherently-precessing simplified model disk should vary as
$r_{out}^{5/2}$, under the assumption that the surface density in the
disk is constant.  To test this predicted dependence, we evolved the
disk over many orbital periods, tracing its growth in size and
corresponding change in precession period; the relationship is
summarized in Figure~\ref{fig:pvsr}.  Note that the disk radius
plotted along the horizontal axis is actually the radius of the
thickest portion of the disk, not the radius of the outermost particle
in the simulation.  The low number density of particles at the outer
disk edge inhibits a consistent and accurate determination of what
actually constitutes the disk size; choosing the radius where the disk
is thickest, on the other hand, provides a more stable and reliable
determination of the disk size.

Starting with an outer radius of $20\, r_S$, and assuming that $a =
0.1$ (see Table 1), the precession period is 69~days.  The period
grows smoothly and monotonically as the disk expands under the action
of viscous angular momentum transport and reaches approximately
600~days when the outer radius is $46\, r_S$.  In this figure, the
solid curve indicates the calculated behavior of the period as a
function of outer disk radius; the dashed curve gives a strict $P
\propto r_{out}^{5/2}$ dependence.  Our simulations clearly exhibit a
period that varies as $r_{out}^{5/2}$; this is a consequence of the
fact that our GC20 and GC30 disks precess as coherent structures.

Future work will including the effect of infalling material and a more
thorough survey of the dependence of disk evolution on black hole
parameters such as $a$.  Very importantly, we will study another
potentially significant signature: the prograde-retrograde flip that
might occur due to sudden changes in the accreted angular momentum
$L$.  Past hydrodynamic simulations \citep{CM97} have hinted that the
complex, clumpy structure of the infalling plasma can lead to
significant fluctuations in both the magnitude and direction of $L$.

\section{Concluding Remarks}\label{sect:conclusion}
We find that the physical conditions associated with accretion onto
Sgr~A* imply the presence of a compact accretion disk that precesses
as a coherent, unwarped structure about the black hole spin axis.  Our
simulations confirm the general results of \citet{NP00}; specifically,
we find that the low-Mach-number disk in Sgr~A* would not be subject
to the Bardeen-Petterson effect.  Given the observed constraints on
disk parameters, we provide an accurate determination of the
precession period of such a disk.  The fact that this period varies as
$r_{out}^{5/2}$ may eventually lead to a compelling determination of
the black hole spin; alternatively, if other independent techniques or
observations can provide accurate estimates of the spin, our results
determine the radius of the disk around Sgr~A*.

Observationally, our work provides some motivation for expecting a
modulation in Sgr A*'s flux with a period of $\sim 50-500$~days; we have
found that the precession period is in this range as long as the disk
size is $\sim 20-30\, r_S$.  Such a modulation might arise in portions
of Sgr~A*'s spectrum produced by an occulted emitter; for example, it
is known that Sgr~A*'s radio emission is produced on scales of
$20-100\, r_S$ (see, e.g., Bower et al. 2004).  Thus, since this disk
is optically thick to cm radiation \citep{LM02b,PM05}, its precession
may lead to a variable aspect that periodically attenuates the total
radio flux from this region.

This effect may already have been observed; \citet{Z01} report a
detection of a $\sim 106$-day cycle in Sgr~A*'s radio emission.  If this
result is eventually confirmed, and if indeed it appears that only a
disk precessing in a non-isotropic gravitational field can account for
it, we may eventually be able to use it as a probe of the spacetime
within the inner $\sim 10\, r_S$ of a Kerr metric and, more
importantly, as a means of directly measuring the black hole's spin.

On the other hand, a non-confirmation of such a modulation on a $\sim 
50-500$-day timescale would tell us that either no disk is present in
Sgr~A* or that its properties---specifically, its size and
orientation---are different from what we now understand, or that the
geometry of the emission region has not yet been identified.  The fact
that such a determination can be made points to the predictive power
of detailed numerical simulations like those reported here and argues
for the need for more sophisticated calculations incorporating the
effects we have ignored thus far.

{\bf Acknowledgments} This work was funded in part under the auspices
of the U.S.\ Dept.\ of Energy, and supported by its contract
W-7405-ENG-36 to Los Alamos National Laboratory, by a DOE SciDAC grant
DE-FC02-01ER41176.  At the University of Arizona, this research was
partially supported by NSF grant AST-0402502, and has made use of
NASA's Astrophysics Data System Abstract Service. F. M. is grateful to
the University of Melbourne for its support (through a Sir Thomas Lyle
Fellowship and a Miegunyah Fellowship).  The simulations were
conducted on the Space Simulator at Los Alamos National Laboratory.

{}

\clearpage

\begin{deluxetable}{lccccc}
\tablewidth{0pt}
\tablecaption{Simulation Properties\label{sims}}
\tablehead{
  \colhead{Simulation}
& \colhead{$i$\tablenotemark{a}}
& \colhead{$a$\tablenotemark{b}}
& \colhead{$\mathcal{M}$\tablenotemark{c}}
& \colhead{$R_{init}$\tablenotemark{d}} 
& \colhead{$N_{init}$\tablenotemark{e}}\\

& \colhead{(deg)}
& 
& 
& \colhead{($r_S$)}
& 

}
\startdata

E1   & 10 & 1\phd\phn   & 12 & 25 & 327357 \\
GC20 & 30 & 0.1 & \phn3 & 20 & 475671 \\
GC30 & 30 & 0.1 & \phn3 & 30 & 757337

\enddata
\tablenotetext{a}{Inclination angle}
\tablenotetext{b}{Black hole spin parameter in units of $GM^2/c$}
\tablenotetext{c}{Mid-plane Mach number}
\tablenotetext{d}{Initial disk outer radius}
\tablenotetext{e}{Initial number of SPH particles}
\end{deluxetable}

\clearpage

\clearpage

\clearpage

\end{document}